\documentclass[aps,prb,showpacs,superscriptaddress,floatfix,twocolumn,notitlepage]{revtex4-2}

\usepackage{blindtext}
\usepackage{pdfsync}
\usepackage{graphicx}
\usepackage{float}
\usepackage[percent]{overpic}
\usepackage{amsmath,amssymb,amsfonts,mathrsfs}
\usepackage{bm}
\usepackage{bbold}
\usepackage{color}
\usepackage{ulem}
\usepackage{natbib}
\usepackage[us,12hr]{datetime}
\usepackage{braket}
\usepackage{bbm}
\usepackage{subcaption}
\usepackage{xcolor}
\usepackage[breaklinks=true,colorlinks,citecolor=blue,linkcolor=blue,urlcolor=blue]{hyperref}

\begin{document}
%\date{Compiled on {\ddmmyyyydate\today} at \currenttime}
\title{Properties of dissipative Floquet Majorana modes using a quantum dot}

\author{Nicol\`o Forcellini}
\email{nforcellini@baqis.ac.cn}
\affiliation{Beijing Academy of Quantum Information Sciences, Beijing 100193, China}
\author{Zhan Cao}
\affiliation{Beijing Academy of Quantum Information Sciences, Beijing 100193, China}
\author{Dong E. Liu}
\affiliation{State Key Laboratory of Low Dimensional Quantum Physics, Department of Physics, Tsinghua University, Beijing, 100084, China}
\affiliation{Beijing Academy of Quantum Information Sciences, Beijing 100193, China}
\affiliation{Frontier Science Center for Quantum Information, Beijing 100084, China}

\date{\today}

\begin{abstract}
We study the electronic conductance of dissipative Floquet Majorana zero modes (FMZMs) in a periodically driven nanowire
coupled to a
quantum dot. We use a numerical method which
can accurately take into account
the dissipation effects from the superconducting
bath, which causes the FMZMs
to have a finite lifetime.
Our results show that, in the weak nanowire-dot
coupling regime, the peak conductance at zero temperature
of the resonant dot
can be well approximated
by a universal function of the FMZM lifetime
rescaled with the nanowire-dot coupling strength:
For a long FMZM's lifetime,
the conductance approaches the characteristic quantized value of $G = e^2/2h$,
whereas $G \rightarrow e^2/h$ (uncoupled dot) as the FMZMs' lifetime goes to zero.
In principle, our method can be used to test the 
presence and lifetime
of FMZMs
in such devices, which is key
for any practical application
of these topological states.
\end{abstract}

\maketitle

%%%%%%%%%%%%%%%%%%%%%%%%%%%%
%%%%%% INTRODUCTION %%%%%%%%%%%%%%
%%%%%%%%%%%%%%%%%%%%%%%%%%%%

\section{Introduction}
Topological superconductors can host zero-energy modes, also called
\textit{Majorana zero modes} (MZMs)~\cite{Read2000,Kitaev2001}, which, if
experimentally engineered, have the potential to make topological quantum computation practical~\cite{kitaev2003,Nayak2008}
due to to their non-Abelian braiding properties~\cite{Nayak2008,leinaas1977,fredenhagen1989,Ivanov2001}.
MZMs have, so far, proven to be extremely difficult to engineer
experimentally. A possible platform is given by semiconductor nanowires with spin-orbit coupling proximitized to an
s-wave superconductor (SC)~\cite{Sau2010,Lutchyn2010,Oreg2010,Alicea2010,Alicea2012}. Seeking the realization of Majorana
modes in such devices has been the subject 
of extensive research in the past decade,
with recent experimental progresses~\cite{Mourik2012,Deng2012,Das2012,Churchill2013,Finck2013,Albrecht2016,Deng2016,ZhangH2017,Gul2018,Zhang2021,Wang2022, Wang2022_2,Song2022,Microsoft2022}.

Topological states of matter
and nontrivial band structures can also
be accessed through Floquet engineering, i.e., the control of quantum systems through
the application of a controlled periodic
drive~\cite{Eckardt2017, Oka2019,rudner2020,Castro2022,Lindner2011,
Kitagawa2011,Dahlhaus2011,Jiang2011,Kitagawa2012,Reynoso2013,LiuPRL2013,Iadecola2013,Fregoso2013,Iadecola2014,FoaTorres2014,Sedrakyan2015,Rechtsman2013,Potter2016,Roy2017,Bomantara2018,Bomantara2018_2,Peng2018}. 
The characteristic ``replicated'' Floquet
band structure in energy space for solid-state systems
has been experimentally verified through time and angle-resolved photoemission spectroscopy~\cite{Wang2013,Mahmood2016,Aeschlimann2021NL}.
In particular, Floquet methods
can induce a topological phase transition
while the static system is topologically
trivial, as for the case 
of Floquet Majorana modes (FMMs), the  periodically-driven 
equivalent of MZMs~\cite{Jiang2011,Reynoso2013,LiuPRL2013}.
The study of non-equilibrium phases of
matter that can exhibit FMMs has been
an active field of research in the past
years, as it connects the field of
topology in condensed matter to problems
in non-equilibrium physics such as 
prethermalization, thermalization and disorder in open/closed quantum
systems, time crystals, etc.~\cite{Mori2016,Abanin2017effective, Abanin2017rigorous, Ponte2015, Sondhi2016,Zhang2017,Choi2017,Bauer2019}.
Specifically, open quantum systems can 
show a complicated behaviour in particle
statistics, depending on the details of the
bath and the system-bath coupling~\cite{Hone2009, Iadecola2015, LiuDong2015_2,Seetharam2015}.
In addition, recent works about
realistic Floquet superconductors~\cite{Qinghong2021} and on
dissipative FMMs in nanowires~\cite{Zhesen}
highlight the importance of taking
the SC bath-nanowire coupling into account:
In the presence of dissipation, while  bosonic condensation in the SC
survives in the presence of a periodic
drive, fermionic quasiparticles,
including FMMs, acquire
a finite lifetime, which the standard Floquet
theorem cannot correctly capture.

\begin{figure}[H]
    \centering
    \includegraphics[width=\linewidth]{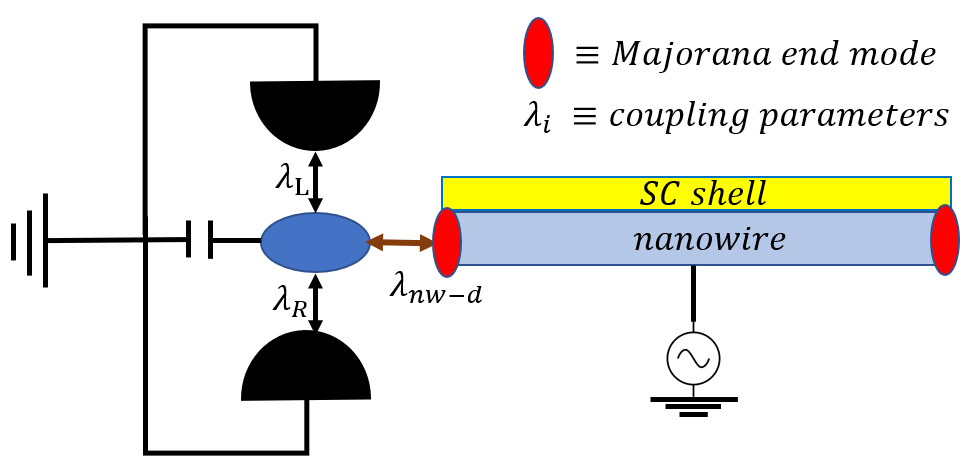}
\caption{Setup of the proposed device: the nanowire is periodically driven with 
frequency $\Omega$, with one of its ends coupled to a quantum dot through an effective coupling $\lambda_{nw-d}$. 
The dot itself is connected to external leads L and R to measure its conductance.}
\label{figspace}
\end{figure}

Therefore, in this work we
further investigate
the lifetime of dissipative
Floquet Majorana zero modes (FMZMs)~\cite{Zhesen}.
The effects of dissipation
from the superconducting bath are taken
into account using the Floquet-Keldysh formalism, which allows to
realistically model
the SC bath
embedding the system~\cite{DongLiu2017,Zhesen,Qinghong2021}.
We model a setup
which would allow to experimentally probe the
lifetime of the dissipative FMZMs, see Fig.~\ref{figspace}.
The setup consists of a quantum dot (QD)
resonantly coupled to a FMZM located at the end of a 
nanowire, which is periodically driven with frequency $\Omega$; two external leads (L,R) are used
to measure the QD electronic conductance.
We show that 
reducing the FMZMs' lifetime
by increasing the periodic
drive amplitude in the nanowire leads to
a transition in the QD
single-spin conductance from $G = e^2/2h$ (Majorana mode signature, see e.g.~\cite{DongLiu2011})
to $G = e^2/h$ (resonant uncoupled dot).

In the weak nanowire-dot
coupling regime, the conductance
 in the presence of dissipation 
 in the nanowire-SC system can be modeled by using a
 finite-lifetime Majorana zero mode (MZM)
 toy model, with
 the conductance being
 a universal function of 
  the FMZM lifetime, rescaled with
  respect to the nanowire-dot coupling
  strength. However, for a stronger coupling
 the Floquet structure of the FMZMs'
 Green's function needs to be taken
 into account: 
 The functional form of the conductance becomes
 more complex and deviates from the simpler
 toy model for short FMZMs' lifetimes, maintaining the
 same asymptotic behaviour for long lifetimes. 

Therefore, we show through our results that the QD conductance
 can be used as a signature of
 the presence of dissipative
 FMZMs in periodically driven topological nanowires,
 and it can also be used as a measure of FMZMs' lifetime
 in such devices.

 The paper is organized as follows:
 In section II we introduce the 
 model of the driven-dissipative nanowire coupled to a 
 QD, and derive the expression of the QD 
 Floquet conductance.
 In section III, we present and 
 discuss our numerical
 results on the QD spectrum and conductance,
 as well as introducing a toy model that 
 allows for a good physical understanding
 of the numerical results. 
 Finally, in section IV, we report our conclusions.

%%%%%%%%%%%%%%%%%%%%%%%%%%%
%%%%%% THEORY %%%%%%%%%%%%%%
%%%%%%%%%%%%%%%%%%%%%%%%%%%%

\section{Model}

\subsection{Superconducting nanowire}
 Consider a
one-dimensional (1D) semiconducting nanowire (SM) in proximity of an
s-wave
superconducting (SC) bath.
We introduce a periodic drive
in the SM region.
The 1D Bogoliubov-de Gennes (BdG) Hamiltonian representing
such a system is given by
\begin{equation}
    H(t) = H_{nw}(t) + H_{sc} + H_c \ ;
\end{equation}
the SM Hamiltonian, in real space, is
\begin{equation}\label{eq:NW_Hamiltonian}
\begin{split}
    H_{nw}(t) = \int_0^L dx \  \psi^{\dagger}_x&\bigg[\left(\frac{p_x^2}{2m}-\mu+ 2A\cos{(\Omega t)} \right)\sigma_0\tau_z \\
    &-\alpha p_x\sigma_y\tau_z
    +V_z\sigma_z\tau_z\bigg] \psi_x,
    \end{split}
\end{equation}
with spinor representation $\psi_x = (c_{x\uparrow },c_{x\downarrow },c^{\dagger}_{x\uparrow},c^{\dagger}_{x\downarrow})^T$,
where $c_{x\uparrow/\downarrow}$ annihilates an up/down-spin
electron
at location $x$ on the nanowire
of length $L$;
$\mu$ is the chemical potential,
$A$ and $\Omega = 2\pi/\tau$ are the amplitude and frequency of the periodic drive,
$V_z$ is the Zeeman energy and $\alpha$ is the spin-orbit coupling strength.
$\sigma_{\mu}$ and $\tau_{\mu}$ indicate
the Pauli matrices in the spin
and Nambu spaces, respectively.
Note that we assume the
nanowire to be uniform, and therefore
effects such as the appearance of
non-topological edge states such as ``quasi-Majorana modes'' (QMMs), which
can be induced by disorder and other imperfections 
in the nanowire~\cite{Kells2012,Moore2018,Moore2018_2,Vuik2019,Donghao2021,Pan2020},
are not considered.
The study of QMMs for this setup goes beyond
the scope of the present work,
and shall be left for a follow-up study.
Moreover, we assume that 
the system is in the nonequilibrium steady-state,
meaning that energy transfer/heating influx and outflux
between
the system and the bath are balanced~\cite{rudner2020,Oka2019,DongLiu2017,Zhesen},
leading to a Green's function (GF) periodic with the
period $\tau$ of the drive $Q(t,t')=Q(t+\tau,t'+\tau)$. 

The Hamiltonian of the SC bath is, in the mean-field BdG form in momentum space
$    H_{sc} = \sum_q \phi^{\dagger}_q \left( \epsilon_q\tau_z
    -\Delta\sigma_y\tau_y \right)\phi_q,
$
where $\Delta$ is the SC gap, and 
$\phi_q = (a_{q\uparrow},a_{q\downarrow},a^{\dagger}_{-q\uparrow},a^{\dagger}_{-q\downarrow})^T$, where
$a_{q\uparrow/\downarrow}$ is the annihilation operator
for spin up/down electrons of momentum $q$ in the SC bath.
The SC-nanowire coupling $H_c$ is modeled as follows:
Firstly, the nanowire Hamiltonian is discretized with lattice spacing
$a$ becoming
   \begin{equation}
   \begin{split}
     {H}_{nw}(t) &= \frac{1}{2}\sum_i \psi^{\dagger}_i
     \big\{
     \left(2t_h -\mu + 2A\cos{\Omega t}\right)\sigma_0\tau_z \\
     &+ V_z\sigma_z\tau_z\big\}\psi_i
     -\left[\psi^{\dagger}_{i+a}\left(t_h\sigma_0 + \frac{\alpha}{2a} \sigma_y\right)\tau_z \psi_i + H.c. \right]
     \end{split}
 \end{equation}
  where the hopping constant $t_h \equiv \hbar^2/2ma^2$.

Then, the Markovian approximation~\cite{Zhesen},
through which
 correlations in the bath are neglected,
allows to couple each site of the chain to independent
and identical SC baths $    H_{sc,i} = \sum_q \phi^{\dagger}_{qi} \left( \epsilon_{qi}\tau_z
    -\Delta\sigma_y\tau_y \right)\phi_{qi}
$.
Hence, the coupling Hamiltonian can be expressed as
\begin{equation}
    H_c =  \sum_{i,q,\sigma} V \left(c^{\dagger}_{i,\sigma}a_{q,\sigma}+a^{\dagger}_{q,\sigma}c_{i,\sigma} \right).
\end{equation}
The SC gap $\Delta$ and the nanowire-SC coupling $V$ are taken
to be real, positive numbers without loss of generality.
The external degrees of freedom can then be integrated
out as shown in~\cite{Zhesen} using the Floquet theorem, and the resulting
effective Floquet Hamiltonian and on-site Green's function
are reported in the Appendix. The main point is that
a finite SC gap $\Delta$ broadens the quasiparticle 
spectrum, representing dissipation caused by the SC bath.
On the other hand, the non-dissipative 
limit $\Delta \rightarrow \infty$
is equivalent to the introduction of a simple 
induced gap term $\Delta_{ind}\sigma_y\tau_y$ in the
nanowire, with induced order parameter $\Delta_{ind}=\pi \rho_F V^2$, where we set $\pi \rho_F = 1$
as the density of states (DOS) of the SC bath. See the Appendix and~\cite{Zhesen} for more details
on the dissipation model
and the large-$\Delta$ non-dissipative limit.
 
\subsection{Coupling the dot}
Consider the undriven QD
coupled to
one of the ends of the
nanowire through some effective hopping $\lambda_{nw-d}$~\cite{DongLiu2011}
\begin{equation}\label{eq:dotcoupling}
\begin{split}
    H_{d} &=\sum_{\sigma} \epsilon 
    _{d} d^{\dagger}_{\sigma}d_{\sigma}
     +\lambda_{nw-d}\sum_{\sigma}
    (d^{\dagger}_{\sigma}c_{L,\sigma}
    + h.c.)
   \\
    &+ \sum_{k,\alpha={L,R},\sigma} 
    \lambda_{\alpha\sigma} (c^{\dagger}_{k\alpha\sigma}d_{\sigma}
    +h.c.)
      + V^d_z(d^{\dagger}_{\uparrow}d_{\uparrow}-d^{\dagger}_{\downarrow}d_{\downarrow}).
\end{split}
\end{equation}
In the above, $\epsilon_d$ is the 
dot level, and $c_L$ is the 
annihilation operator
for the last site of the nanowire.
The two leads
labeled as left (L) and right (R)
have Hamiltonian
$H_{leads} = \sum_{k,\alpha = {L,R}}
c^{\dagger}_{k\alpha\sigma}c_{k\alpha\sigma}$ and couple to the dot 
with a width
$\Gamma_{\alpha\sigma}\equiv 2\pi |\lambda_{\alpha\sigma}|^2\rho_{Fl}$,
where the lead DOS $2\pi\rho_{Fl} = 1$ is assumed
to be constant.
$c_{k\alpha\sigma}(d_\sigma)$ denotes the electron annihilation
operator for the leads (QD).
$V_z^d$ is the Zeeman splitting
for the dot, which might be different
from $V_z$ of the nanowire, or, when 
assuming that the QD and nanowire 
materials are the same, can be set
to have the same value. 
In any case, we assume that the 
Zeeman energy is the largest scale in
the QD, and therefore we can consider only one spin channel, ignoring any electron-electron interaction in the above Hamiltonian.
Moreover, we tune the QD such that
the energy of the spin-$\downarrow$ electron 
$\epsilon_d -V_z^d = 0$, allowing this state
to be resonant with the FMZM.

From this model, and following the
method described in~\cite{Zhesen},
we use the recursive Floquet-Green's function technique
to obtain the relevant components of the dot Green's function used for our calculations --see the Appendix for more details.
 
In this work we investigate 
the QD conductance due to
its coupling to dissipative FMZMs
of finite lifetime $\tau_{FM}$:
As mentioned in the previous section, a finite SC gap $\Delta$
induces dissipation, i.e., a broadening of the quasiparticle spectrum.
Our operational definition of the
FMZM lifetime is the
inverse of the width $\Gamma_{FM} = \tau_{FM}^{-1}$ obtained by fitting
the FMZM peak of the spectrum at the end 
of the nanowire:
With the (time-averaged)
DOS
of a FMZM at the end
of the nanowire
\begin{equation}
    \nu_{FM}(\omega)
    = -\frac{1}{\pi}\text{Im}{\text{Tr}\left\{Q^R_{FM}(0,\omega)\right\}},
\end{equation}
and the FMZM Floquet Green's function
having the form $Q^R_{FM}(0,\omega)=[\omega-\Sigma_{FM}(\omega)]^{-1}$,
then one has, in the zero-frequency approximation,
\begin{equation}
    \nu_{FM}(\omega)
     \propto \frac{\Gamma_{FM}}{\omega^2+\Gamma_{FM}^2}.
\end{equation}
Hence, we identify the self-energy $\Sigma_{FM}(\omega= 0) = -i\Im{[\Sigma_{FM}(\omega = 0)]} =  i\Gamma_{FM}$
in the FMZM Green's function as purely imaginary, since 
the peak of $\nu_{FM}(\omega)$ is always at $\omega = 0$
for a FMZM. 
 In addition, increasing the driving
 amplitude $A$ also leads to an increase
 in dissipation and a decay in quasiparticle
 lifetime in the nanowire~\cite{Zhesen}.
 Hence, we use $A$ as the tuning parameter
 for FMZMs' lifetime control.
For a more detailed discussion on the definition of
Floquet Majoranas' lifetime,
we refer to~\cite{Zhesen}. For this work, the above definition suffices.

\subsection{Quantum dot conductance for a Floquet system}
 The time-dependent current in the
 left lead is given by 
 \begin{equation}
    I_L(t) = \frac{ie}{\hbar}\sum_{k\sigma} 
    (\lambda_{kL\sigma}c^{\dagger}_{kL\sigma}d_{\sigma} - \lambda^*_{kL\sigma}d^{\dagger}_{\sigma}c_{kL\sigma}),
\end{equation}
  with the equivalent definition for the current through R. 
Details of the derivation of the expression
for the current $I(t) = I_L(t) + I_R(t)$
through the QD using Floquet-Keldysh field theory
are left to the Appendix. Here, we report the final
expression for the zero-bias time-averaged conductance ($G = d\langle I \rangle / dV|_{V\rightarrow 0}$)
\begin{equation}\label{eq:G}
    G = -\frac{2e^2}{\hbar}\int \frac{d\omega}{2\pi} \frac{\Gamma_L\Gamma_R}{\Gamma_L+\Gamma_R}\text{Im}\text{Tr}\left\{{Q^R_{dd}}(0,\omega)\right\}\left(-\frac{\partial n_F}{\partial \omega}\right).
\end{equation}
  In the above, $n_F(\omega)$ is the Fermi-Dirac distribution, and $Q^{R}_{dd}(0,\omega)$
is the 0$^{\text{th}}$ Fourier component (time average) of the
retarded component of the QD GF defined as --with $t_{rel} = t-t'$--
\begin{equation}\label{eq:Floquet_GF_Fourier}
    Q(n,\omega)
    = \frac{1}{\tau}
    \int^{\tau}dt \int dt_{rel} e^{-in\Omega t} e^{-i\omega t_{rel}}Q(t,t')
\end{equation}
for $n=0$, where we made use of the periodicity of the GF~\cite{DongLiu2017}.
Since in our system we assume that the driving frequency $\Omega$ is very high compared to the other energy scales,
it is sufficient to compute the time-averaged conductance. 
Eq.~\ref{eq:G} can be seen as the Floquet generalization of the static
QD conductance~\cite{DongLiu2011}.

As a special case, consider symmetric
coupling to the leads $\Gamma_L = \Gamma_R = \Gamma$ leading to the peak conductance (as the temperature $T\rightarrow 0$)
\begin{equation}\label{eq:peakG}
    G_{peak} = -\frac{e^2}{h} \Gamma \ 
    \text{Im}\text{Tr}\left\{Q^R_{dd}(0,\omega\rightarrow 0)\right\}.
    \end{equation}
For our numerical calculations
of the conductance,
we only select the 
spin
$\downarrow$-electron.
From now on, $G_{peak}$ will be referred to 
as $G$, and Eq.~\ref{eq:peakG} will be used to
produce our numerical results for the time-averaged
QD conductance.

\section{Results and discussion}

\subsection{Quantum dot spectral function}

\begin{figure}[H]
\centering
\subfloat[]{
	\label{subfig:dot_spectrum}
	\includegraphics[width=0.45\textwidth]{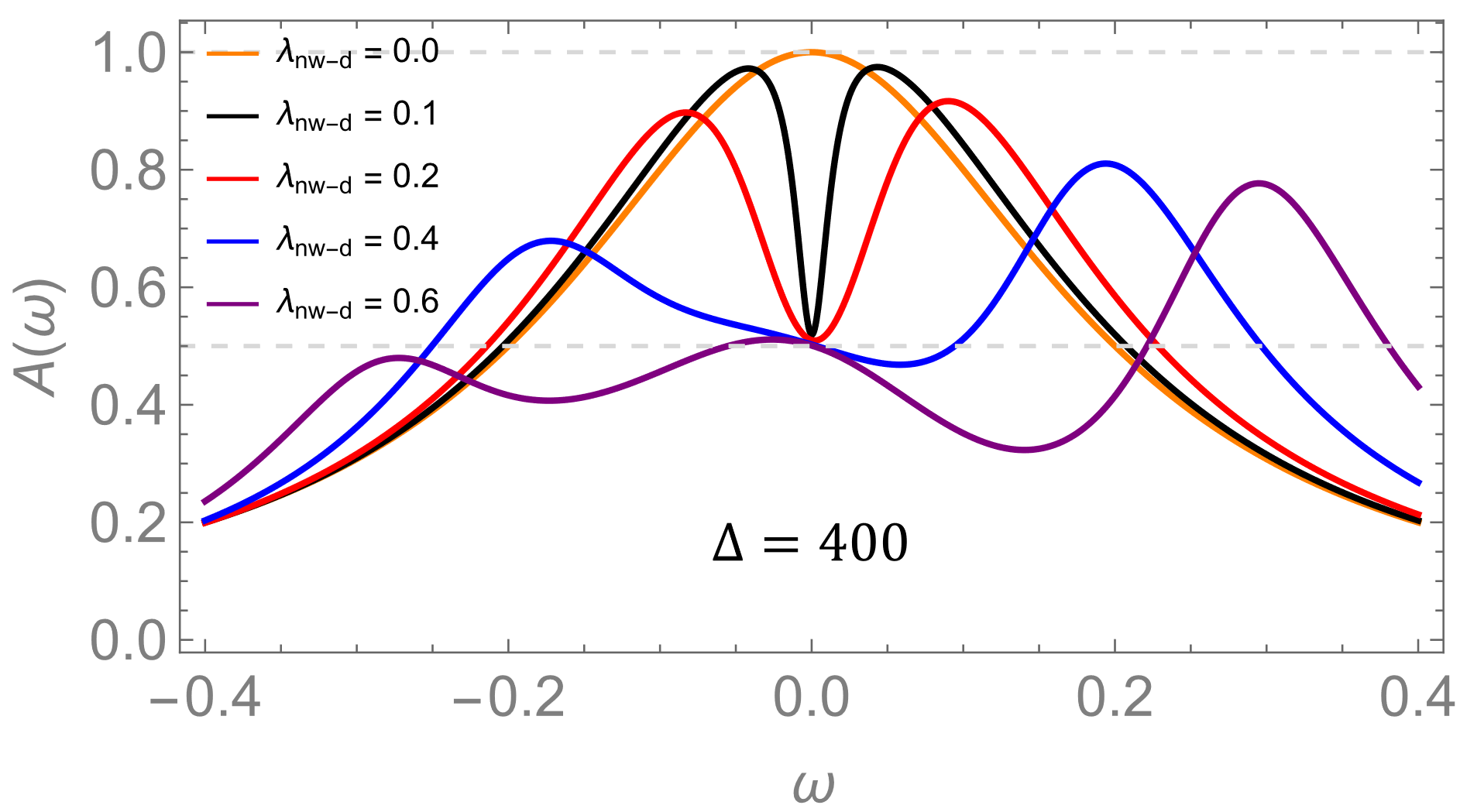} }
	\hfill
\subfloat[]{
	\label{subfig:spectrum_dissipative}
	\includegraphics[width=0.45\textwidth]{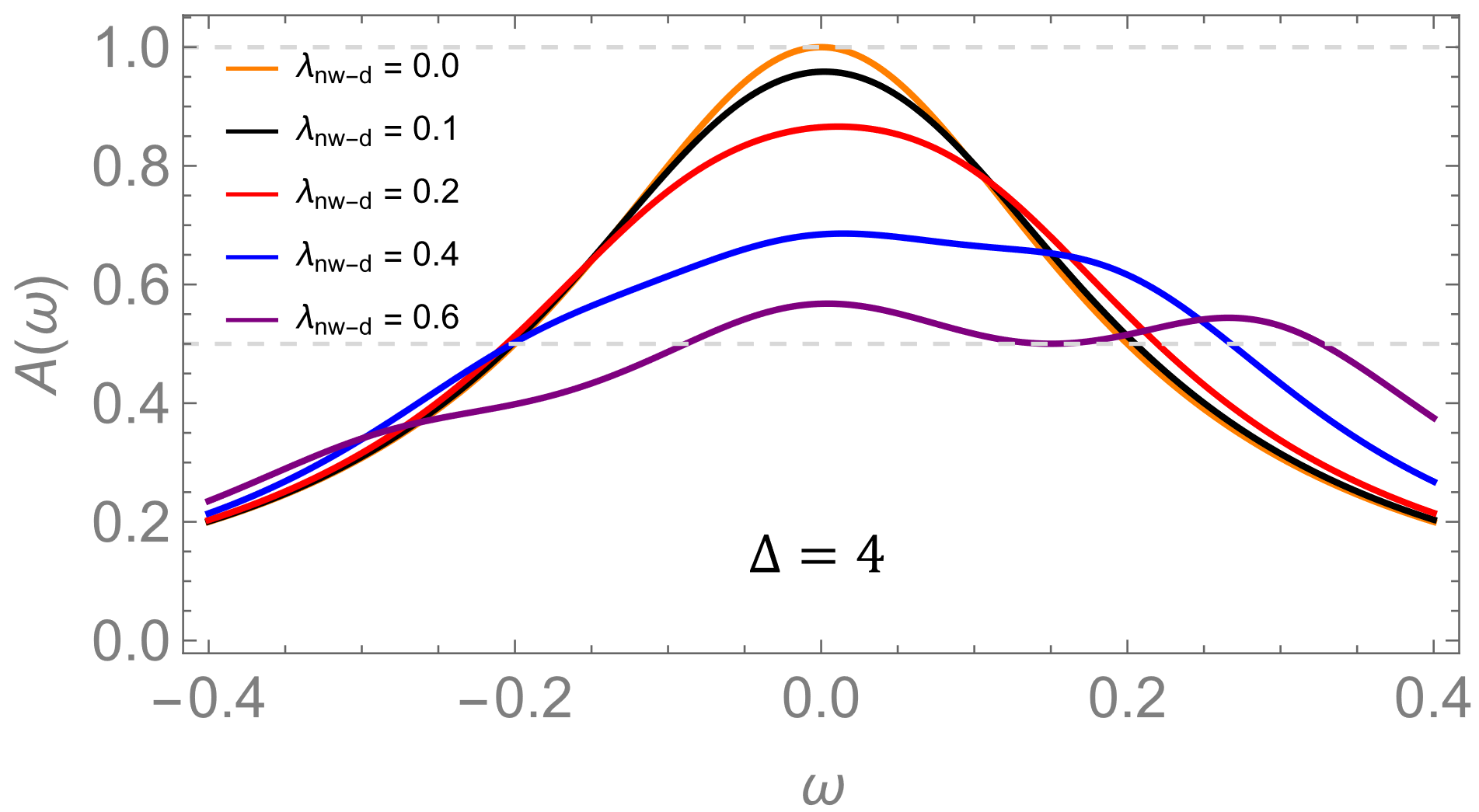} } 
\caption{Quantum dot spectrum when coupled to (a) a non-dissipative FMZM,
note the value $A(0) = 1/2$, expected from Majorana modes for any $\lambda_{nw-d} \neq 0$, and (b) a dissipative FMZM, with the same parameters as in (a) except for the
SC Gap $\Delta$. Note how the $1/2$-value at $\omega = 0$ is lost, with $A(0) \rightarrow 1$ for small $\lambda_{nw-d}$. 
For the calculations, $\epsilon_d = V^d_z$, placing the $\downarrow$-electron 
at zero energy. $\Gamma_L=\Gamma_R=\Gamma = 0.1$,
$t_h = 2.0$ (bandwidth $D=4t_h = 8.0$), $\mu=-2.0$, $\alpha = 3.0$, $V_z =V_z^d= 4.0$,
$V=1.2$, $A=3.0$ and $\Omega = 12.0$.}
\label{fig:spectrum_dot}
\end{figure} 

As a first step, in Fig.~\ref{fig:spectrum_dot} we investigate
 the single-spin spectral function 
of the dot $A_{\downarrow}(\omega) \equiv - \Gamma \Im\left\{Q^R_{dd,\downarrow}(0,\omega)\right\}$ --for simplicity we will drop the $\downarrow$-label.
In Fig.~\ref{subfig:dot_spectrum}, in the
non-dissipative limit,
the dot spectrum has the same characteristics of a quantum dot coupled to a (non-Floquet,
static) Majorana zero mode (MZM)
with the same setup~\cite{DongLiu2011}:
The dot spectral function 
$A(0) = 1/2$
whenever the MZM is coupled to the dot, independently of the coupling strength, and $A(0) = 1$ with
$\lambda_{nw-d}=0$ (resonant isolated dot).
This translates in
a peak conductance with a value 
$G = e^2/2h$, which is a signature of
MZMs~\cite{DongLiu2011}.
This is in fact not surprising: Expressing a
FMZMs as a Bogoliubov 
quasiparticle operator
\begin{equation}\label{eq:Floquet_Majorana}
    \eta^{\dagger}(x,t)
    = \int dx \left[ u(x,t)\hat{\psi}^\dagger_x + v(x,t)\psi_x \right],
\end{equation}
it
has been proved, both numerically and using
theoretical arguments, that
FMZMs are MZMs 
at all times~\cite{LiuPRL2013,Reynoso2013,Shtanko2020}. In the
above expression the
functions $u(x,t)$ and $v(x,t)$
represent the spatial
and temporal
modulation of the zero modes,
which, albeit time-dependent, are localised 
at the end of the nanowire at all times. 

On the other hand, Fig.~\ref{subfig:spectrum_dissipative} shows the
spectrum in the presence of a dissipative
FMZM: In this case, the $A(0) =1/2$-signature
is generally lost, with a decreasing $\lambda_{nw-d}$
leading to $A(0)\rightarrow 1$.
Intuitively, this is due to the FMZM having acquired
a finite lifetime due to dissipation,
which approaches the uncoupled limit 
$A(0)\rightarrow 1$ as the lifetime gets
shorter and/or the coupling $\lambda_{nw-d}$ 
gets weaker.
$\lambda_{nw-d}$ should be kept small enough: with the current parameters --see caption of Fig.~\ref{fig:spectrum_dot}-- the
induced gap parameter is $\Delta_{ind} = 1.44$,
and therefore we must have $\lambda_{nw-d} < \Delta_{ind}$
to limit the undesired influence of modes above the gap.
In our calculations,
the nanowire has a length $L=200$ 
sites, and we set the Floquet cutoff
$N_F = 10$ to ensure convergence
of the numerics~\cite{Zhesen}.
The parameters used in this work and
reported in Fig.~\ref{fig:spectrum_dot} and~\ref{fig:toymodel_MZMf} agree with previous works on
dissipative and non-dissipative
FMZMs
\cite{Zhesen,LiuPRL2013},
and are consistent with proposals of FMZM realization in cold-atom systems \cite{Jiang2011}.

\subsection{Conductance and FMZMs' lifetime}

\begin{figure}[H]
\centering
	\includegraphics[width=0.47\textwidth]{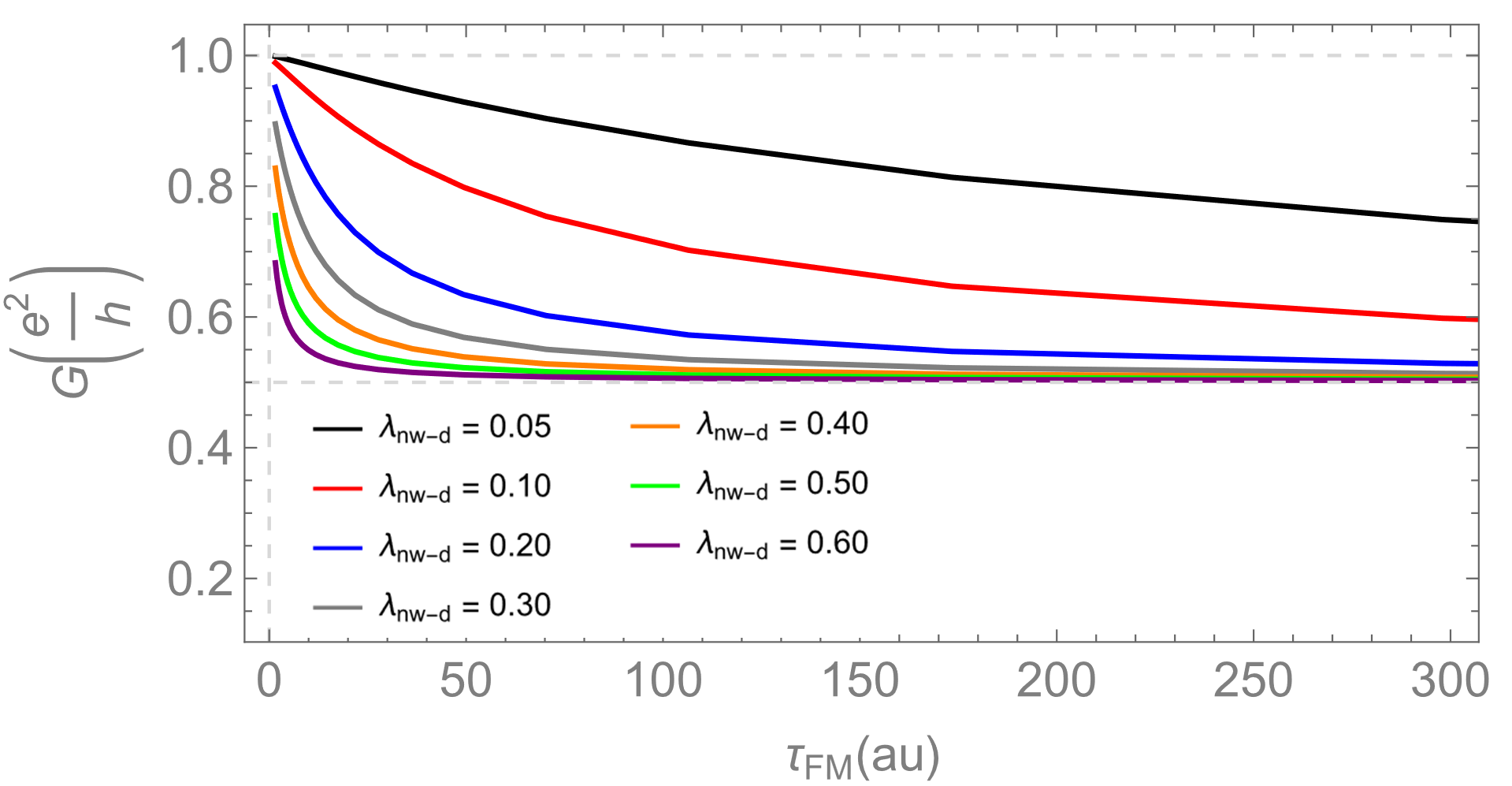}
\caption{Conductance $G$ from Eq.~\ref{eq:peakG}
as a function of the FMZM lifetime $\tau_{FM}$ (arbitrary units), as the nanowire-dot
coupling $\lambda_{nw-d}$ is
changed.
With a sufficiently long lifetime,
the conductance $G$
is stabilized around the
topological signature $G=e^2/2h$. 
A larger value for $\lambda_{nw-d}$, meaning
a stronger coupling to the finite-lifetime FMZM,
leads to a more stable signal around $G=e^2/2h$.
Parameters as in 
Fig.~\ref{fig:spectrum_dot} ($\Omega = 12.0$, etc.),
with amplitudes $0 \leq A \leq 5.2$, which allows to tune the lifetime $\tau_{FM}$ in the x-axis, and $N_F=10$ Floquet matrix cutoff.}
\label{fig:conductance}
\end{figure}

In order to better understand the 
conductance signal of the dissipative FMZMs, we
show in Fig.~\ref{fig:conductance}
the values of $G$ as a function of 
FMZM lifetime $\tau_{FM}$ for different
$\lambda_{nw-d}$.
By increasing the strength $A$ of
the Floquet drive (and shortening the FMZM lifetime
$\tau_{FM}$) we find
a transition between
$G=e^2/2h$, indicating
the presence of a Floquet
Majorana mode, to $G=e^2/h$,
which is the conductance of
an uncoupled dot resonant at zero energy. 
A stronger coupling stabilizes
the conductance signal around the 
Majorana signature $G=e^2/2h$.
In the following, we start from
the simplest toy model for dissipative Majorana modes
in order to better analyze the numerical results. 

\subsubsection{Toy model of a dissipative Majorana mode}

Consider the following 
toy model for a normal MZM coupled to the dot.
The effective Hamiltonian for the single-spin dot-MZM system is
\begin{equation}\label{eq:MZMtoymodel_hamiltionian}
    H = H_{leads}+H_T+H_{QD-MZM},
\end{equation}
where
\begin{equation}
    H_{QD-MZM} = \epsilon_d d^{\dagger}d + \lambda_{nw-d}(d-d^{\dagger})\eta_1 + i\delta\eta_1\eta_2,
\end{equation}
with $\delta$ being the exponentially small coupling $\delta \sim e^{-L/\xi}$ with coherence length $\xi = v_F/\Delta_{ind}$. Here, we set $\delta = 0$
to isolate one of the Majoranas.
$H_{leads}$ and $H_T$ are the single-spin versions of the lead and lead-dot coupling Hamiltonians previously defined after Eq.~\ref{eq:dotcoupling}.
In order to impose a finite lifetime for $\eta_1$,
we assign it a width $\Gamma_{M}(\omega)$
in its GF $Q^R_{\eta}(\omega) = C_0/[\omega + i\Gamma_{M}(\omega)]$, where $C_0$ is a normalization constant.

Assuming symmetric coupling to the leads $\Gamma_L=\Gamma_R=\Gamma$ and
$\epsilon_d = 0$, the retarded component $Q^R_{dd}(\omega)$ for this toy model is given by
\begin{equation}\label{eq:toymodel_GF}
    Q^R_{dd}(\omega)=\frac{\left[1-\frac{C_0|\lambda_{nw-d}|^2}{(\omega+i\Gamma_M(\omega))(\omega+2i\Gamma)}\right]}{(\omega+2i\Gamma)\left[1-\frac{2C_0|\lambda_{nw-d}|^2}{(\omega+i\Gamma_M(\omega))(\omega+2i\Gamma)}\right]}.
\end{equation}
This leads to $G$ with the following simple form
\begin{equation}\label{eq:toymodel_G}
    G(\tilde{\lambda}) = \frac{e^2}{h}\frac{1+C_0 \tilde{|\lambda|^2}/2\Gamma}{1+C_0\tilde{|\lambda|^2}/\Gamma},
\end{equation}
where $|\tilde{\lambda}|^2 = |\lambda_{nw-d}|^2/\Gamma_M(\omega \rightarrow 0) \sim |\lambda_{nw-d}|^2\tau_M$.
Hence, we found that, given a certain $\Gamma$
coupling strength with the leads, $G$ is a universal function of the
rescaled coupling $\tilde{\lambda}$.
This coupling
represents the two competing energy/timescales of the nanowire-QD
system: MZM lifetime $\tau_{M}$ and the inverse of the nanowire-QD coupling strength $\sim |\lambda_{nw-d}|^2$.

We can easily check that 
the function has the expected behaviour,
with the main features of the conductance curves that can be seen by inspection of Eq.~\ref{eq:toymodel_G}:
For a perfect MZM with $\tau_M \rightarrow \infty$, $G_{peak} = e^2/h$ if the QD is decoupled 
from the MZM, while $G_{peak}=e^2/2h$
for any finite $\lambda_{nw-d}$.
For a finite $\tau_M$, $G_{peak} > e^2/2h$. Specifically, reducing 
$\tau_M$ would smoothly increase
the peak conductance to $e^2/h$.
Moreover, a larger $\lambda_{nw-d}$ brings $G_{peak}$ closer to
the quantized value $e^2/2h$, 
just as observed in Fig.~\ref{fig:conductance}.
However, since Eq.~\ref{eq:toymodel_G} can only
model MZMs, in the next section we modify the
toy model to include the Floquet structure.

 \subsubsection{Floquet Green's function correction}

  \begin{figure}[H]
\centering{
	\includegraphics[width=0.45\textwidth]{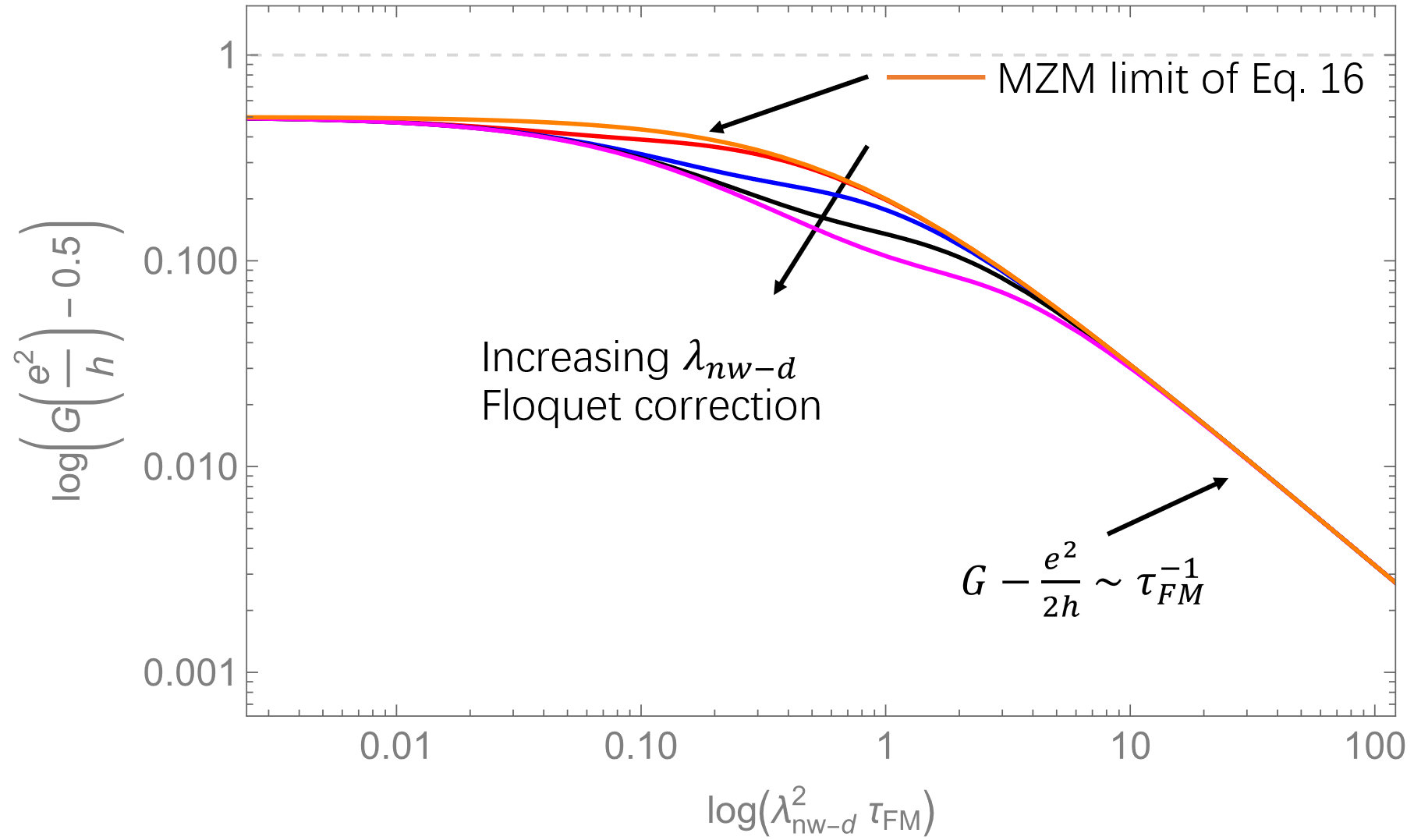} }
\caption{Log-Log plot of the conductance $G -e^2/2h$ from Eq.~\ref{eq:G_Floquet} using our toy model corrected for the Floquet
structure of the FMZM Green's function, Eq.~\ref{eq:Floquet_GF}.
In this plot, $C_k \sim k^{-\alpha}$, $\alpha = 2$.
The figure shows that, with increasing $\lambda_{nw-d}$,
the curve deviates from the low-$\lambda_{nw-d}$ limit of 
Eq.~\ref{eq:toymodel_G}, which is shown in orange,
only for small enough lifetimes. The values set for $\lambda_{nw-d}$ are $1.0$ (red line, almost overlapping with the orange line),
$2.0$ (blue) and $3.0$ (black)
and $4.0$ (magenta).
The values are
chosen arbitrarily in order to show the
qualitative behavior due to the Floquet structure,
with $\Omega = 12.0$.
}
\label{fig:conductance_toymodel_Floquet}
\end{figure}
 
 Due to their time-periodicity, the FMZMs
 of Eq.~\ref{eq:Floquet_Majorana} can be expanded as
 $\eta(t) = \sum_n e^{-in\Omega t} \ \eta_n$, and
 since their quasienergy $\epsilon_{\alpha} = 0$,
 the Green's function is given by
\begin{equation}\label{eq:Floquet_GF}
    Q^R_{\eta}(n,\omega) = \sum_{k = -\infty}^{\infty} \frac{\eta_{k+n}(\eta_{k})^{*}}{\omega-k\Omega+i\Gamma_M(\omega)},
\end{equation}
where the $\Gamma_M(\omega)$ must be the same for each $k$ due to the Floquet theorem,
as shown in~\cite{Zhesen}.  By definition, we have $\eta^{\dagger}(t) = \eta(t)$,
and therefore $(\eta_n)^* = \eta_{-n}$. Hence, for $n=0$ the Green's
function can be written as
\begin{equation}
     Q^R_{\eta}(0,\omega) = \frac{C_0}{\omega+i\Gamma_M(\omega)} + \sum_{k=1}^{\infty} \frac{2 C_{k}(\omega + i \Gamma_M(\omega))}{(\omega+i\Gamma_M(\omega))^2-(k\Omega)^2},
\end{equation}
where $C_k \equiv \eta_{k}(\eta_{k})^{*}$, so that $C_k\geq 0$. 
These $k$-dependent factors can be considered as decaying as a function of $k$, due to the properties of the Fourier series.
For our numerical calculations, we set
$C_k \sim k^{-\alpha}$, $\alpha = 2$.
Different choices for the form
of the decaying $C_k$ lead to qualitatively similar results.

The dot Green's function becomes
\begin{equation}\label{eq:toymodel_FGF}
    Q^R_{dd}(\omega)=\frac{\left[1-\frac{|\lambda_{nw-d}|^2}{(\omega+2i\Gamma)}Q^R_{\eta}(0,\omega)\right]}{(\omega+2i\Gamma)\left[1-\frac{2|\lambda_{nw-d}|^2}{(\omega+2i\Gamma)}Q
    ^R_{\eta}(0,\omega)\right]}.
\end{equation}
From this expression,
the generalization of Eq.~\ref{eq:toymodel_G} is easily obtained as
\begin{equation}\label{eq:G_Floquet}
    G(\tilde{\lambda}) = \frac{e^2}{h}\frac{1+{|\tilde{\lambda}|^2}/2\Gamma\left(A^0+\sum_{k=1}^{\infty}\frac{2A^k\Gamma^2_M}{\Gamma_M^2+(k\Omega)^2}\right)}{1+{|\tilde{\lambda}|^2}/\Gamma\left(A^0+\sum_{k=1}^{\infty}\frac{2A^k\Gamma^2_M}{\Gamma_M^2+(k\Omega)^2}\right)},
\end{equation}
where $\Gamma_M \equiv \Gamma_M(\omega \rightarrow 0)$.
Note that the introduction of the energy scale defined by $\hbar\Omega$
does not allow $G$ to be a universal function of the dimensionless
parameter $\tilde{\lambda}$,
since the expression
in brackets has a dependence on the FMZM width $\Gamma_M$. In the $\Omega\rightarrow \infty$-limit,
the expression reduces to the previous model.
We can also check the relevant limits of Eq.~\ref{eq:G_Floquet} in the same way that we did
for Eq.~\ref{eq:toymodel_G}: For the uncoupled dot, 
$G=e^2/h$, and the large-$\tau_{FM}$
limit ($\Gamma_M\rightarrow 0$)
leads to $G = e^2/2h$, with the
Floquet contribution (second term in the
bracket) vanishing.
This leads to a $G(\tilde{\lambda})$
as reported in Fig.~\ref{fig:conductance_toymodel_Floquet}.
The results show a deviation from the 
form of Eq.~\ref{eq:toymodel_G}
for shorter lifetimes, while maintaining
the same $G-1/2 \sim \tau_{FM}^{-1}$
behavior for long lifetimes, in agreement with our numerical results. \\

To summarize, in this section we showed that the conductance
through a resonant quantum dot as a function of the
rescaled coupling/lifetime $|\lambda_{nw-d}|^2\tau_{FM}$
has a characteristic functional form, which is
almost identical when the dot is coupled to either
dissipative
MZMs or FMZMs, with the conductance curves
differing only for short lifetimes
and strong enough nanowire-dot coupling. 
In particular,
given the possibility of tuning of the FMZMs' lifetime
and the nanowire-dot coupling strength,
our results provide
a signature for the presence
and stability --in terms of
their lifetimes-- of FMZMs in topological
nanowires: The QD conductance should
behave as shown in Figs~\ref{fig:conductance} and \ref{fig:conductance_toymodel_Floquet}
and as described by Eq.s~\ref{eq:toymodel_G} and \ref{eq:GF_MZMf}. An assumption that 
we make in the model is the one of a resonant
dot: If the QD energy level $\epsilon_d \neq 0$,
then the $G(\tau_{FM}\rightarrow 0) < e^2/h$. It is in principle
always possible to tune the QD in such a way by measuring its conductance
when uncoupled from the nanowire.

In the next 
and final section, we briefly explore how the QD conductance
is modified if other modes in the nanowire also couple 
to the QD; this can happen in non-ideal experimental conditions.

\subsection{Soft SC gap: normal fermion coupling}

\begin{figure}[H]
\centering
\subfloat[]{
	\label{subfig:conductance_complete}
	\includegraphics[width=0.45\textwidth]{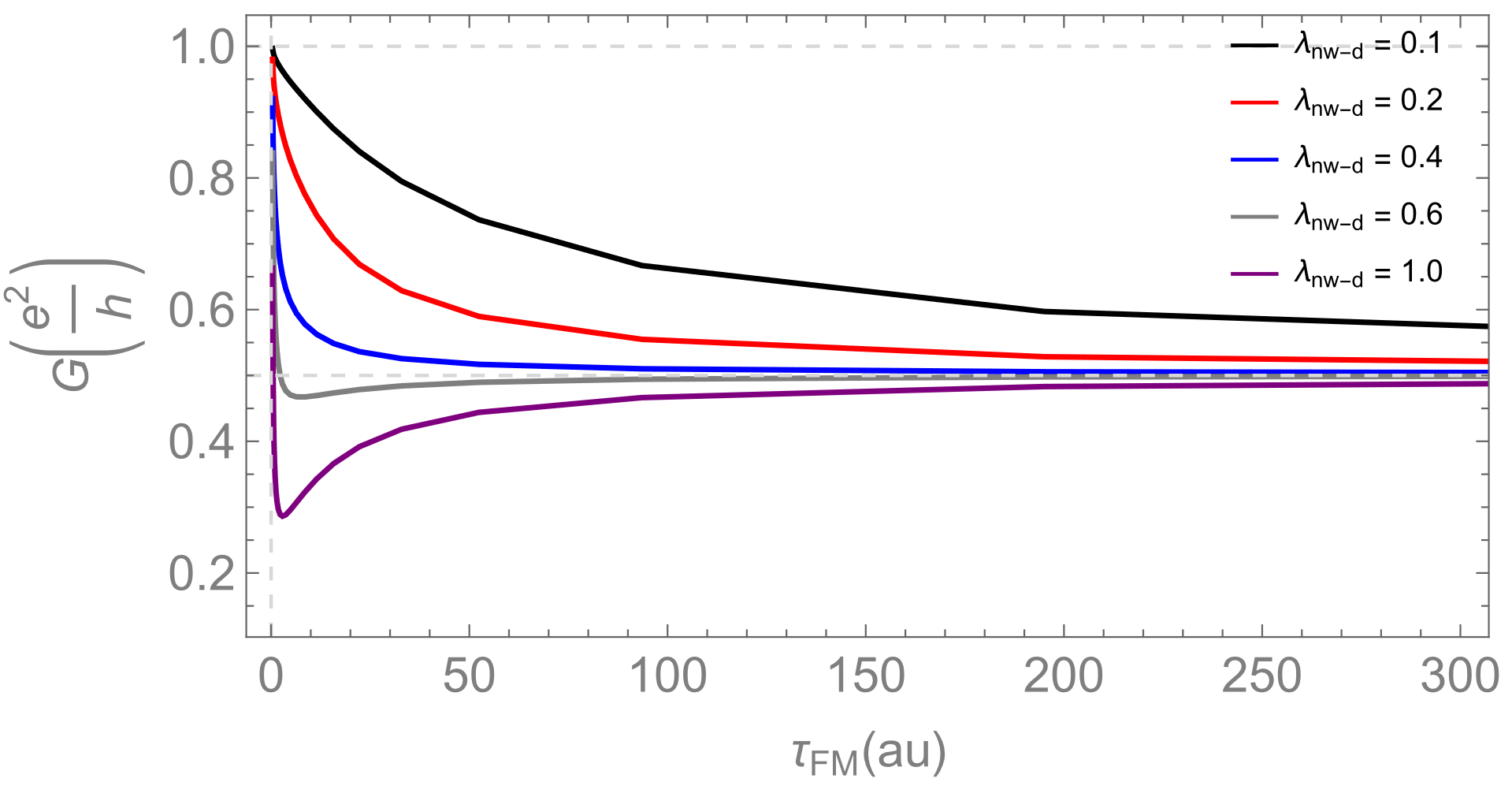} }
	\hfill
\subfloat[]{
	\label{subfig:conductance_complete_toymodel}
	\includegraphics[width=0.45\textwidth]{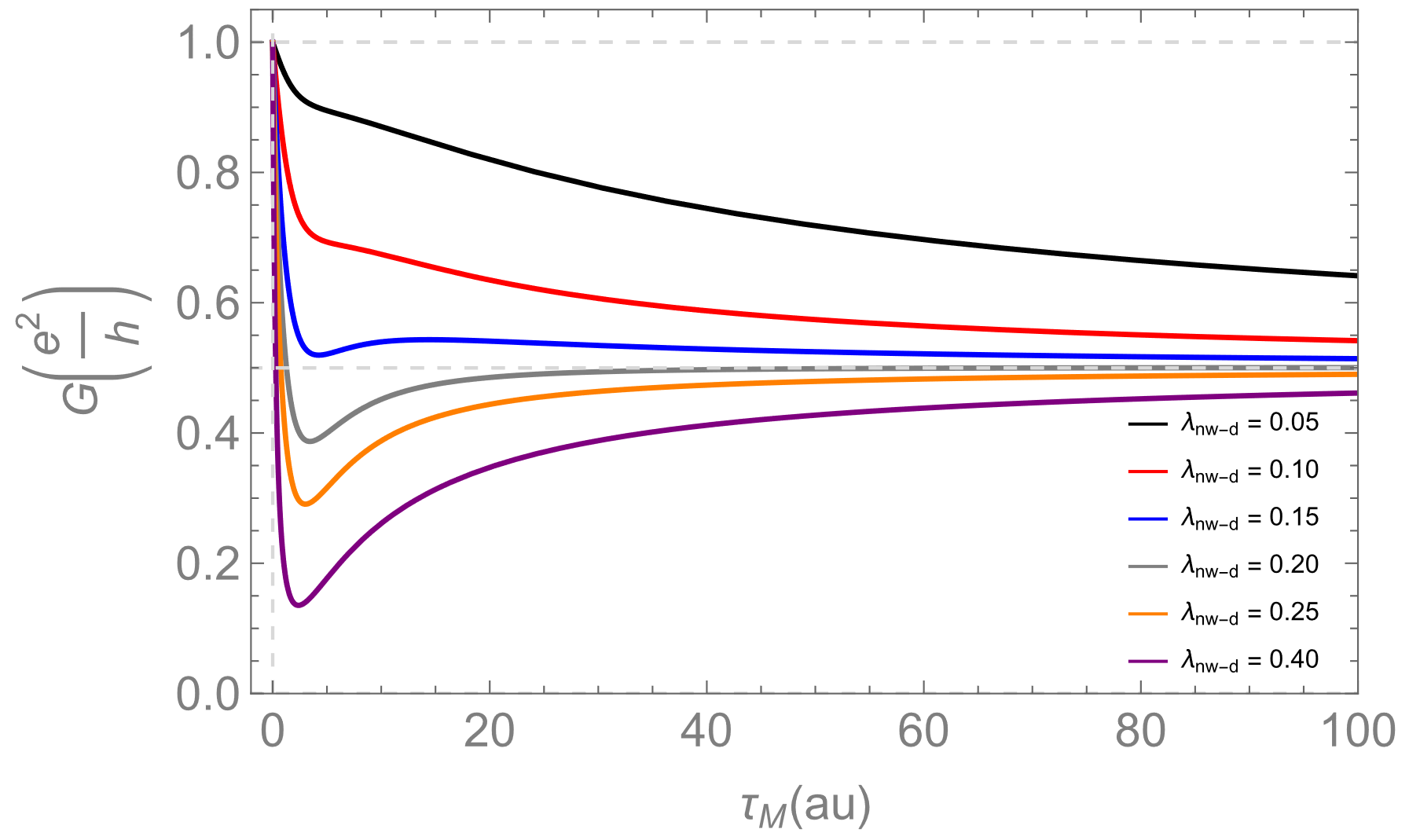}
	}
\caption{ (a) Same plot as Fig.~\ref{fig:conductance}
showing $G<e^2/2h$ for short-$\tau_{FM}$ and 
large $\lambda_{nw-d}$, where the model parameters 
have been modified to get FMZMs with smaller
$\Delta_{ind}=0.64$. Specifically,
$t_h = 1.0$ ($D=4.0$), $\alpha = 1.5$, $V_z =V_z^d= 1.2$,
$V=0.8$, $\Omega = 6.0$, and $0 \leq A \leq 5.2$ as in 
Fig.~\ref{fig:conductance},
with a Floquet matrix cutoff $N_F = 20$.
(b) $G(\tau_M)$ from Eq.~\ref{eq:conductance_MZMf} for
the toy model with an additional
off-resonant fermion:
we can see that the model can qualitatively reproduce the
large-$\lambda$ short-lifetime behavior shown in (a).
 Eq.~\ref{eq:conductance_MZMf} is computed with $\Gamma = 0.1$, $\delta = 0.3$.}
\label{fig:toymodel_MZMf}
\end{figure}

A setting in which the undesired coupling could happen
is the case in which
the nanowire-dot coupling gets too large, the coupling of
states above the induced gap can affect the value of the conductance.
For instance, this effect can manifest itself
when engineering
very short FMZMs' lifetimes, corresponding to
larger amplitudes in the drive, as explained in section II.B 
and as established in~\cite{Zhesen}: in such a case,
the spectrum
in the nanowire is broadened and the SC gap
derived from the
Floquet GF becomes soft~\cite{Qinghong2021}.
The effect on the conductance is shown in Fig. \ref{subfig:conductance_complete}, where
we modify our model parameters and we set some $\lambda_{nw-d}\gtrsim \Delta_{ind}$:
the numerical results show values $G<e^2/2h$ when
 approaching shorter FMZM lifetimes.
However, these conductance curves show a minimum
for small $\tau_{FM}$,
 reverting
to $G \rightarrow e^2/h$ for
$\tau_{FM} \rightarrow 0$.
A value of $G = 0$ is a signature
of a dot coupled to a normal fermion: Hence,
the features seen in the figure are probably due to
the contribution of states above the induced gap,
detected via the strong coupling.
However, since such quasiparticles also acquire
a finite lifetime due to the drive and dissipation,
the limit $G(\tau\rightarrow 0) = e^2/h$ should be maintained.

To analyze this behavior,
we add to the MZM a normal fermion with finite
width $\Gamma_f(\omega)$ and with finite energy $\delta$
--slightly off-resonant with the QD--
to the MZM toy model. Hence, the GF
becomes --in particle-hole space--

\begin{equation}\label{eq:GF_MZMf}
    Q^R_{nw}(\omega) \approx \begin{pmatrix}
\frac{1}{\omega + i\Gamma_{M}} + \frac{1}{\omega-\delta + i\Gamma_{f}} & 0   \\
0 & \frac{1}{\omega + i\Gamma_{M}} + \frac{1}{\omega+\delta + i\Gamma_{f}}  \\
\end{pmatrix}.
\end{equation} 
In order to derive the simplest possible expression
for $G$ capturing the numerics, we make the following 
assumptions: The lifetimes
of both the MZM and the off-resonant fermion are set to be the same
$\Gamma_f=\Gamma_M\sim 1/\tau_M$, as
well as the nanowire-QD coupling $\lambda_{nw-d}$.
This leads to the following expression
for the conductance (always with $\epsilon_d = 0
$):
\begin{equation}\label{eq:conductance_MZMf}
\begin{split}
    G(\tau_M) \!=\! \frac{e^2}{h}\frac{1+\delta^2|\lambda_{nw-d}|^2\tau_M/[2\Gamma(\delta^2+1/\tau_M^2)]
    }{1\!+\!|\lambda_{nw-d}|^2\!\left[\frac{\tau_M}{\Gamma}\!+\!\frac{1}{(\delta^2+1/\tau_M^2)}\!\left(\!\frac{1}{\Gamma\tau_M}\!+\!\frac{|\lambda_{nw-d}|^2}{2\Gamma^2}\!\right)\!\right]}.
\end{split}    
\end{equation}

A plot of the above function is shown in
Fig.~\ref{subfig:conductance_complete_toymodel}:
The model is able to replicate the 
small-$\tau_M$/large-$\lambda$ features of the numerical simulation of~\ref{subfig:conductance_complete}, which we can
now explain as follows:
On the one hand, the FMZM's lifetime is
reduced by dissipation, leading to
a reduced ``effective coupling''
$\tilde{\lambda}$ to the dot.
On the other hand, the effect of SC gap
softening in the nanowire as the amplitude of the
periodic drive is increased
can lead to a signal $G\rightarrow 0$
as the dot starts coupling to a normal
fermion above the gap. This effect only
becomes important as the nanowire-QD
coupling $\lambda_{nw-dot}\gtrsim\Delta_{ind}$.

\section{Conclusions}
In this work, we studied for the first time the transport signatures of dissipative FMZMs coupled to a resonant quantum dot. We derived an expression for the conductance from first principles via the Floquet-Keldysh formalism, allowing for a nonperturbative treatment. We showed that the conductance of a dot coupled to a FMZM shows a characteristic transition from the Majorana signature $G=e^2/2h$ to the uncoupled value $G=e^2/h$ as the FMZM lifetime decays. We showed that the conductance can be well approximated by a universal function of the FMZM lifetime, rescaled with respect to the nanowire-dot coupling strength, despite the fact that the true functional form of the conductance is in fact more complex due to the Floquet structure of the FMZMs' GF. Indeed, we showed that the Floquet correction only becomes important at short lifetimes and strong nanowire-dot coupling.
 
Periodically driven 
nanowires seem to be a convenient
platform for the study of the lifetime
and stability of FMZMs, given the
simple nanowire-QD coupling setup
and tunability of the periodic drive.
However, the setup requires a good degree of 
control and fine-tuning of the coupling strength 
with the QD, as well as making sure
that the induced superconducting gap stays 
large enough under a strong periodic drive,
which might represent obstacles for an 
experimental realization.
Nonetheless, with the above premises, our work
illustrates a clear signature for the 
presence of FMZMs
in topological nanowires, even when taking into account 
the effects of dissipation from the SC bath.
In future work, it would be interesting to include 
more realistic effects in the setup, such as 
disorder and imperfections in the 
periodically driven dissipative nanowire
to study their effects on electronic transport.

\section{ACKNOWLEDGMENTS}
We thank Gu Zhang, Zhesen Yang and Qinghong Yang for helpful discussions. The work has been supported by the National Natural Science Foundation of China  (Grants No.~11974198 and No.~12004040), Beijing Academy of Quantum Information Sciences, Beijing 100193, China.

\appendix
\section{}

\subsection{Nanowire model and recursion method}
After integrating out the bath degrees of freedom, 
the on-site retarded component of the nanowire Floquet
Green's function (GF) takes 
the form 
 \begin{equation}\label{eq:retGF}
      \underline{g_i(\omega)} = 
      \left[ \underline{\omega}-\underline{{H}_{eff,i}(\omega)} \right]^{-1}.
  \end{equation}
where the on-site effective Floquet Hamiltonian is
  \begin{widetext}
  \begin{equation}\label{eq:Floquet_hamiltonian}
      \underline{{H}_{eff,i}(\omega)} = 
      \begin{pmatrix}
... &  & & &  \\
 & {H}_{nw,i}-\Omega +\Sigma_{sc}(\omega + \Omega) & A\sigma_0\tau_z &0 &   \\
 & A\sigma_0\tau_z  & {H}_{nw,i} +\Sigma_{sc}(\omega) & A\sigma_0\tau_z &  \\
   & 0 & A\sigma_0\tau_z & {H}_{nw,i}+\Omega +\Sigma_{sc}(\omega - \Omega) &    \\
  &  &  & & ...

\end{pmatrix},
  \end{equation}
 \end{widetext}
 where $H_{nw,i} = (2t_h-\mu)\sigma_0\tau_z+V_z\sigma_z\tau_z$
and the bath self-energy term is~\cite{Zhesen}
 \begin{equation}\label{eq:self_energy}
     \Sigma_{sc}(\omega) = V^2
     \frac{1}{\sqrt{-(\omega+i\eta)^2+\Delta^2}}
     \left[-(\omega+i\eta)-\Delta\sigma_y\tau_y\right],
 \end{equation}
 where $\eta = 0^+$ and the bath
 DoS is assumed to be uniform.
 For the numerical calculations $\eta$ is set to a finite positive value,
 much smaller than any other energy scale in the system.
The notation
 $\underline{M}$ indicates a matrix
 in Floquet space,
  which is in principle infinite-dimensional
  due to the Fourier expansion, and
  we denote by $[\underline{M}]_{mn}$.
  
  For instance, in Eq.~\ref{eq:Floquet_hamiltonian}
  $  [\underline{{H}_{eff,i}(\omega)}]_{nn} =  {H}_{nw,i} +\Sigma_{sc}(\omega-n\Omega)+n\Omega$.
   In $ \underline{{H}_{eff,i}(\omega)}$
 the off-diagonal
  elements represent the harmonic drive. A value of $\kappa \equiv A/\Omega < 1$
ensures convergence and allows for the truncation of
the matrices in the Floquet Hilbert space for any
value of $\Omega$, and the matrix dimensions can be
kept conveniently small without
any appreciable loss of accuracy~\cite{DongLiu2017,Zhesen}.
Moreover, the $Q(0,\omega)$-elements of the 
  Floquet GF used in the main text to compute time-averages
  of observables are extracted from the 
  $[\underline{Q_{dd}(\omega)}]_{00}$-component of the Floquet GF matrix.
The meaning of the self-energy of Eq.~\ref{eq:self_energy} is that it represents dissipation through its $\omega$-dependence,
i.e., a broadening of the quasiparticle
spectrum via its imaginary part;
when $\Delta < \Omega$,
 the self-energy of 
 $[\underline{{H}_{eff,i}(\omega)}]_{11/-1-1}$ becomes purely imaginary, which
 means that ``single-photon'' Floquet transitions lead to energy-particle 
 exchange directly above the SC bath; when $\Delta > \Omega$, higher-order 
 transitions are necessary and therefore the FMZM lifetime
 is longer.
The non-dissipative limit is found by
letting $\Delta \rightarrow \infty$, whence
$\Sigma_{sc} = \Delta_{ind}\sigma_y\tau_y$,
with the self-energy simply becoming a real-valued
induced gap parameter in the nanowire, with $\Delta_{ind}\equiv \rho_F V^2$.
 
 The spectrum and LDoS of the FMZMs at the end of the nanowire
 can be calculated from the retarded part of
 the local GF, which is found by using the following recursive matrix equation
 in the Floquet-Keldysh-BdG space~\cite{Zhesen}
 \begin{equation}\label{eq:recursive}
     \underline{Q_{i+1,i+1}}(\omega)
     = \big[\underline{g}_i^{-1}(\omega) -\underline{T_{i+1,i}}
     \cdot \underline{Q_{i,i}}(\omega) \cdot \underline{T_{i,i+1}}\big]^{-1},
 \end{equation}
 where $\underline{g}_i(\omega)$ is the on-site ``bare'' GF
 defined in Eq.~\ref{eq:retGF}. The above equation is
 iterated for $N=200$ sites in our calculations,
 with an appropriate $N_F$ Floquet matrix
 cutoff to ensure convergence.
 The hopping matrix in the nanowire is
 \begin{equation}
     T_{i,i+1} = {T_{i+1,i}}^T
     = 
\begin{pmatrix}
-t_h & 0 & -\alpha/2a & 0 \\
0 & t_h & 0 & \alpha/2a \\
 \alpha/2a & 0 & -t_h & 0  \\
0 & -\alpha/2a & 0&t_h
\end{pmatrix}
 \end{equation}
 which, extended in F-K-BdG space,
is simply $\underline{T_{i,i+1}}= I_{2N_F+1} \otimes I_{2} \otimes T_{i,i+1}$.
 
The quantum dot coupled to 
the leads of Eq.~\ref{eq:dotcoupling} can be 
included as an additional 
site of the recursive chain, 
which means that we need to perform
an additional iteration of Eq.~\ref{eq:recursive}
with $g^{-1}(\omega) = \text{diag}[\omega -\epsilon_d-V_z^d+i(\Gamma_L+\Gamma_R),\omega -\epsilon_d+V_z^d+i(\Gamma_L+\Gamma_R),\omega +\epsilon_d+V_z^d+i(\Gamma_L+\Gamma_R),\omega +\epsilon_d-V_z^d+i(\Gamma_L+\Gamma_R)]$
and $ T_{i,i+1} = \text{diag}[-\lambda_{nw-d},\lambda_{nw-d},-\lambda_{nw-d},\lambda_{nw-d}]$.

  \subsection{Derivation of the current and conductance using Floquet-Keldysh field theory}

In order to derive the expression of
conductance of
Eq.~\ref{eq:G} in the main text,
one can start from
 the effective action with current source term~\cite{DongLiu2015}
\begin{equation}
      S = S_0 + S_{L-D} +S_{source},
\end{equation}
where
\begin{equation}
\begin{split}
    S_{0} = 
    \sum_{kk',\alpha = L,R}
    & \int_C\int_C dtdt' \Psi_{k,\alpha}^{\dagger}(t)
    Q^{-1}_{0,kk'\alpha}(t,t')\Psi_{k',\alpha}(t') \\
    &+ \int_C\int_C dtdt' \Psi_{d}^{\dagger}(t)
    Q^{-1}_{0,dd}(t,t')\Psi_{d}(t'),
\end{split}
\end{equation}
and the coupling part of the action is given by
\begin{equation}
\begin{split}
    S_{L-D} &=
    \sum_{k\alpha} \int_C dt (\lambda_{k\alpha }c^{\dagger}_{k\alpha}d+h.c.) \\
    &= \sum_{k\alpha} \int_C dt
    (\Psi^{\dagger}_{k,\alpha}(t)
    \hat{M}_{T,k\alpha}\Psi_d(t)+h.c.),
\end{split}
\end{equation}
where we choose to work in the Nambu basis
$\Psi_{k,\alpha}^{\dagger} = (c^{\dagger}_{k,\alpha},c_{k,\alpha})/\sqrt{2}$ and $\Psi_{d}^{\dagger} = (d^{\dagger},d)/\sqrt{2}$;
the tunnelling matrix element
is 
$
  M_{T,k\alpha} =   
  \begin{pmatrix}
  \lambda_{\alpha k} & 0 \\
  0 & -\lambda^*_{\alpha k}
  \end{pmatrix};
$
$Q_{0,dd}(t,t')$ is the
Green's operator for the dot,
and $Q_{0,kk'\alpha}(t,t')$
is the Green's function
of the lead. 

By defining a spinor for the
whole space 
$\Psi^{\dagger} = 
(l^{\dagger}_{kL},l_{kL},d^{\dagger},d, l^{\dagger}_{kR},l_{kR})/\sqrt{2}$, the above action terms
can be expressed as
\begin{equation}
    S_0 + S_{L-D} = 
    \int_C\int_C dtdt'
    \Psi^{\dagger}(t)
    Q^{-1}(t,t')\Psi(t'),
\end{equation}
where the Green's function is
\begin{equation}
    Q_{kk'} = 
    \begin{pmatrix}
  {Q_{Lk,Lk'}} & {Q_{Lk,d}} & {Q_{Lk,Rk'}} \\
  {Q_{d,Lk'}} & {Q_{d,d}} & {Q_{d,Rk'}} \\
  {Q_{Rk,Lk'}} & {Q_{Rk,d}} & {Q_{Rk,Rk'}} 
  \end{pmatrix}.
\end{equation}
Out of these components,
for the transport calculations,
we only need the $Q_{dd}$ from
the dot, as we will show
in the following derivation.

Finally, the source term
is defined as
\begin{equation}
    S_{source} = -\int dt A(t)I_L(t)
    = -\sum_{a,b=1}^2\int_{-\infty}^{\infty}dt \Bar{\Psi}_a \hat{A}_{ab}\hat{M}_L\Psi_b,
\end{equation}
where the spinors and matrices are now
in Keldysh space after 
performing a Larkin-Ovchinnikov
rotation, where 
$\hat{A}=A^q\gamma^q$, $\gamma^q=\sigma_1$,
and the Keldysh spinors $\Psi_{1,2}$, $\Bar{\Psi}_{1,2}$
are defined as
$\Psi_{1/2} = (\Psi^+\pm\Psi^-)/\sqrt{2}$
and $\bar{\Psi}_{1/2} = (\bar{\Psi}^+\mp\bar{\Psi}^-)/\sqrt{2}$~\cite{kamenev_2011,altland_simons_2010}.
This source term generates the current through the 
left lead
 \begin{equation}
    I_L(t) = \frac{ie}{\hbar}\sum_{k\sigma} 
    (\lambda_{Lk}l^{\dagger}_{Lk\sigma}d_{\sigma} - \lambda^*_{Lk\sigma}d^{\dagger}_{\sigma}l_{Lk\sigma})
    = \vec{\Psi}^{\dagger}(t)\hat{M}_L\vec{\Psi}(t),
\end{equation}
The transport matrix 
$\hat{M}_L$ is defined
as
\begin{equation}
    \hat{M}_L = \frac{ie}{\hbar}
  {  \begin{pmatrix}
  0 & {M^{12}_{L}} & 0 \\
  {M^{21}_{L}}& 0 & 0 \\
  0 &   0 & 0 
  \end{pmatrix}},
\end{equation}
with $M^{12}_{L} = 
\scriptsize{\begin{pmatrix}
  \lambda_{Lk} & 0  \\
  0 & \lambda^*_{Lk} 
  \end{pmatrix}}$
  and $M^{21}_{L} = 
\scriptsize{\begin{pmatrix}
  -\lambda^*_{Lk} & 0  \\
  0 & -\lambda_{Lk} 
  \end{pmatrix}}$.
The generating function is
$
    Z[A] = \int D[\bar{\Psi}\Psi]
    e^{iS}
$,
and upon Gaussian integration to linear order in $A^q$,
$
    \ln{Z[A]} = \text{Tr}\ln{[\hat{1}-QAM]}
    \approx -\text{Tr}[QA^q\gamma^q M_L]
$. The current can be expressed as 
\begin{equation}
      I_L(t) =
    \frac{i}{2}
    \frac{\delta \ln{Z[A]}}{\delta A^q}\bigg|_{A^q=0} \approx -\frac{i}{2}\text{Tr}[Q(t,t)\gamma^q M_L],
\end{equation}
leading to
\begin{equation}
\label{eq:IL_Floquet}
\begin{split}
     I_L(t)  
    =\frac{e}{2\hbar}\sum_k\sum_n\int & \frac{d\omega}{2\pi} e^{in\Omega t} 
    \text{Tr}[Q^K_{Lk,d}(n,\omega)M^{21}_L \\
    &+Q^K_{d,Lk}(n,\omega)M^{12}_L],
\end{split}
    \end{equation}
where the trace over the lead-QD space has been
performed, and the following identity was applied:
\begin{equation}
    \text{Tr}[Q_{\alpha}\gamma^q] = Q^K_{\alpha}(t,t) = \sum_n  \int \frac{d\omega}{2\pi} e^{in\Omega t} 
    Q^K(n,\omega),
\end{equation}
where $Q^K$ is the
Keldysh component of
the Green's function.
The above Green's functions (for the left lead) can be expressed
as follows, in terms of the lead
GF $g^0_{Lk}$ and dot GF $Q_{dd}$:
\begin{equation}
     {Q_{d,Lk}} = {M_T^{21}}
     {Q_{dd}}\cdot   {g^0_{Lk}},
\end{equation}
\begin{equation}
\begin{split}
     {Q_{Lk,d}} &= 
 {M_T^{12}}{g^0_{Lk}} 
     \cdot {Q_{dd}}.
\end{split}
\end{equation}
 
Taking the Keldysh component of these
products leads to
\begin{equation}
     {(Q_{d,Lk}})^K = {M_T^{21}}
     [({Q_{dd}})^R({g^0_{Lk}})^K + ({Q_{dd}})^K({g^0_{Lk}})^A ],
\end{equation}
and 
\begin{equation}
     ({Q_{Lk,d}})^K = {M_T^{12}}
     [({g^0_{Lk}})^R({Q_{dd}})^K + ({g^0_{Lk}})^K({Q_{dd}})^A ],
\end{equation}
where $(Q,g^0)^{R/A}$
are the QD/lead Green's function
retarded and advanced
components.
Upon substitution in Eq.~\ref{eq:IL_Floquet}, 
it leads to
the following expression
for the time-dependent current
\begin{equation}\label{eq:IL_complete}
    \begin{split}
 I_L&(t)
    =
    \frac{ie}{2\hbar}\sum_ne^{in\Omega t }\int \frac{d\omega}{2\pi} 
   \Gamma_L \{(1-2n_L(\omega))
   \\
    &\times[{Q_{dd}}^R(n,\omega)-{Q_{dd}}^A(n,\omega)] 
    - {Q_{dd}}^K(n,\omega)
    \},
\end{split}
\end{equation}
where $n_L(\omega)$ is 
the Fermi-Dirac distribution
of the L-lead.
Note that, at this stage, the expression shows the
exact current with its full time-dependence. The only assumption, as stated in the main text, is that the system
is in a nonequilibrium steady-state, and thus
the Green's function is periodic in time with
the period $\tau$ of the drive.
In addition,
for the derivation of Eq.~\ref{eq:IL_complete}, the following are used:

\begin{itemize}
\item{Identities for the lead GF, $(g^0_{Lk}(\omega))^K = -2\pi i \delta(\omega-\epsilon_k)[1-2n_L(\omega)]$ and $(g^0_{Lk}(\omega))^R-(g^0_{Lk}(\omega))^A = -2\pi i \delta(\omega-\epsilon_k)$.}
\item{The summation over $k$ is performed with the help of the $\delta$-function, and
we assume the wide-band limit for the leads, with a constant density of states $\rho(\omega)=\rho_{Fl}$.}
    \item{The line-width function is defined as
    $\Gamma_L =
    2\pi\rho_{Fl}|\lambda_L|^2$.}
\end{itemize}

The equivalent expression
can also be derived for $I_R(t)$, defined as

\begin{equation}
    I_R(t) = \frac{ie}{\hbar}\sum_{k\sigma} 
    (\lambda_{Rk}l^{\dagger}_{Rk\sigma}d_{\sigma} - \lambda^*_{Rk\sigma}d^{\dagger}_{\sigma}l_{Rk\sigma}).
\end{equation}

For a time-dependent system, $I_L(t) = - I_R(t)$ only holds for time averages, i.e. $\langle I_L(t) \rangle =-\langle I_R(t) \rangle $~\cite{haug2008quantum}. 

Therefore, for
the time-averaged current 
through the dot
$\langle I \rangle = \langle (I_L-I_R)/2 \rangle $ the following simple
expression for the current is valid:
\begin{equation}\label{eq:I_complete}
\begin{split}
    \langle I \rangle &= 
    \frac{ie}{\hbar}
    \int \frac{d\omega}{2\pi}
    [n^L_F(\omega)-n^R_F(\omega)] \\
    &\times
    \text{Tr}\left\{\frac{\Gamma_L\Gamma_R}{\Gamma_L+\Gamma_R}[Q^R_{dd}(0,\omega)-Q^A_{dd}(0,\omega)] \right\}.
\end{split}
\end{equation}

The above leads directly to the expression for the
conductance of Eq.~\ref{eq:G} in the main text.

%%%%%%%%%%%%%%%%%%%%%%%%%%%%%%%%%%%%%
%%%%%%%%% BIBLIOGRAPHY %%%%%%%%%%%%%%
%%%%%%%%%%%%%%%%%%%%%%%%%%%%%%%%%%%%%
\bibliographystyle{apsrev4-1} 
%\bibliography{ref} 

\end{document}